\documentstyle[12pt]{article}
\begin{document}
\begin{center}
{\Large  Quantum Cosmology and the value of $\Lambda$ }\\

\bigskip
{\bf D.H. Coule}\\
Department of Applied Mathematics\\
University of Cape Town \\
7700 Rondebosch, RSA.
\bigskip
\begin{abstract}
We analyse a simple quantum cosmological model with just a  $\Lambda$
term present.
 Differing  results are
obtained depending on the boundary conditions applied. In the
Euclidean regime  the Hartle-Hawking boundary condition
gives the factor $\exp(1/\Lambda)$ but, in agreement with
Rubakov et al.[10], is badly behaved for
negative $\Lambda$.  Tunneling boundary conditions
suggest an initially large value for $\Lambda$.

 If only
Lorentzian regions are considered all boundary conditions suggest an initially
large value of $\Lambda$ for spatial
curvature $k=1$. This differs from the previously
obtained result of Strominger [11] for such models.

\end{abstract}
\end{center}
\newpage
{\bf Introduction}\\
One possible solution to
the cosmological constant $\Lambda$ problem
that  has attracted a lot of interest is due
to the idea that wormhole solutions can lead $\Lambda$ to become a
dynamical variable with a distribution function $P(\Lambda)$ [1]
- for a review of this proposal see eg. ref.[2]. It is
suggested that this
function is peaked, due to De Sitter instantons, with  the Baum-Hawking
factor $P(\Lambda)\sim exp(1/\Lambda)$ [3,4] so predicting $\Lambda
\rightarrow 0$ [1].

 Wormholes are used in two distinct ways in such arguments,
firstly they are used to justify why $\Lambda$ should be treated as a
dynamical quantum variable instead of a usual classical variable. This
is the most important aspect, as it allows one to make
 predictions  of the possible values of $\Lambda$.
 Wormholes have further
been used in connecting many universes together which produces a further
exponentiation [1] i.e.
\begin{equation}
P(\Lambda) \sim \exp\left (\exp \left ( 1/\Lambda \right )
\right )
\end{equation}
This is only useful if the first factor $\sim 1/\Lambda$ is correct. In
other words, this aspect of wormholes only exaggerates any underlying
behaviour.
In this letter we wish to analyse a quantum cosmological model to see
if the Baum-Hawking factors are present or justified. Although this
simple model has been studied by many authors there are widely
differing predictions for the expected value of $\Lambda$.
\\

{\bf Cosmological constant model}

 When only a  cosmological constant is present the Wheeler-DeWitt
(WDW) equation
 takes the  form (ignoring factor ordering corrections) see eg.[5-7]
\begin{equation}
\left ( \frac{d}{da^2}-U \right )\Psi(a)=0
\end{equation}
Where
the WDW potential $U$ for  a closed $k=1$ universe is given by
\begin{equation}
U=a^2-\Lambda a^4
\end{equation}

The potential is sketched in Fig.(1). This has been studied by
many authors especially as the case of quantum tunnelling to a
Lorentzian universe [8,9]. We follow particularly the analysis of
Rubakov  et.al. [10] and Strominger[11].
The WKB solutions have the form cf.[10]
\begin{equation}
\Psi=\frac{1}{\sqrt{|U(a)|}}exp\left (\pm\int \sqrt{U(a)}da\right )
\end{equation}
where the `action' $S=-\int\sqrt{U}da$ is given by
\begin{equation}
-\int a(1-\Lambda a^2 )^{1/2}da=\frac{(1-\Lambda a^2)^{3/2}}{3\Lambda}
\;\;.
\end{equation}
Taking the limits between $a=\Lambda^{-1/2}$
and  $a=0$  gives the solutions
\begin{equation}
\Psi_{\pm}\sim \exp(\pm 1/\Lambda).
\end{equation}
The (+) sign corresponds to the Hartle-Hawking (HH)[12]
 boundary condition $\exp(-S)$
and the (-) sign to the tunnelling one $\exp(-|S|)$- see
for example [5-7].

 If we assume that the probability
of having a specific $\Lambda$ is $P(\Lambda)\sim \Psi^2 \sim \exp (\pm 2
/\Lambda )$
 then the two approaches predict $\Lambda \rightarrow 0$ and
$\Lambda \rightarrow \infty$ respectively.
For the HH case we appear to get the suppression of $\Lambda$ but the
opposite for the tunnelling case. However the tunnelling occurs through
the barrier to $U=0$ where $a^2\sim 1/\Lambda$. When the barrier is
small i.e.  when $\Lambda $ is
large tunnelling is enhanced: this is somewhat
analogous to the application of an electric potential to an atom which
allows electron  to tunnel away. In this case a large value of $\Lambda$
is enhancing the possibility of the universe tunnelling into existence.

According to Strominger [11] because the scale factor $a$ today is very large
the value of $\Lambda\sim a^{-2}$ is very small as required to fit
observation. This does not however agree with the usual interpretation
of quantum cosmology which is that of predicting initial conditions. As
quantum  tunneling is expected to occur to an initial size of roughly
Planck dimensions the initial value of $\Lambda$ is correspondingly
big $\sim 1$ in Planck units. If instead the initial scale factor
was large (and so
the initial value of $\Lambda
\sim$ small) it would mean that the Euclidean domain would extend to
large sizes. It would  then be inconsistent with the present structure
of space-time which appears Lorentzian down to at least sizes of $\sim
10^{-20}$ meters. To restate this point: quantum cosmology
should not compute conditional
probabilities for classical epochs of the universe
long after the quantum era is over. So in contrast to ref.[11], the
question ``given that the scale factor is $\sim 10^{60}$ what is the
probability distribution for $\Lambda$ '' is not the correct question to
ask of the WDW equation. Rather we should be asking: what is the
distribution function for $\Lambda$ when the classical epoch of the
universe first started ?
\\

We now consider  a problem that has arisen for the case of a
$-ve$ cosmological constant [10]. There seem
a number of unnecessary complications in this analysis,
such as 3rd quantization and addition of matter fields, that
can be neglected, but with the problem (of $\Lambda \rightarrow
-\infty$) still remaining. We therefore repeat their arguments in
a more simplified and transparent manner.
The equivalent expression to
eq. (5) for negative $\Lambda$  is
\begin{equation}
S= -\frac{(1+|\Lambda|a^2)^{3/2}}{3|\Lambda |}
\end{equation}
 It is no longer clear
what integration  limits have to be placed on $a$. Choosing them from
$a=0$ to $a$ gives the solutions

\begin{equation}
\Psi\sim exp\pm \left[\frac{(1+|\Lambda | a^2)^{3/2}}{3|\Lambda |}
-\frac{1}{3|\Lambda|}\right ]
\end{equation}
If we keep track of the signs then the (+) one corresponds to the HH case
and will be dominated by large
\begin{equation}
\sim \left [ \frac{(1+|\Lambda|a^2)^{3/2}}{
3|\Lambda |}-\frac{1}{3\Lambda }\right ]
\end{equation}
 i.e. by $\sqrt{|\Lambda |}a^3$ large.\footnote{
If we had not subtracted the part corresponding to $a=0$ we would
also find a divergence when $|\Lambda|\rightarrow 0$}.
 This is the problem that the HH boundary condition predicts
$\Lambda\rightarrow -\infty$
found by  Rubakov et. al.[10]
\footnote{It was not necessary, as done in ref.[10] to include matter fields
or to consider a third quantized theory to obtain this
dominant $\sqrt{|\Lambda|}a^3$ factor.}
 It is uncertain that this makes any sense and
is rather an artifact of the HH wavefunction being peaked around the
exponentially increasing solution. cf. Fig.(2) in Ref.[9].\\
There is another reason to discount this solution. If the cosmological
constant was absent the wave function would be\\
\begin{equation}
\Psi \sim \exp \left ( \pm a^2/2 \right )\;\;.
\end{equation}

If we choose the ($+$)  sign there is  a contradiction with
our notions of classical behaviour since the universe would apparently
prefer to have large size. Rather the other sign is more correctly
peaked around $a=0$. This point has recently been made by Vilenkin [13]
in criticism of the ``generic'' boundary conditions, but which is
also valid against the HH ones when $\Lambda$ is negative.\\
 The other (-) sign  solution in eq.(8)  formally appears to
predict $\Lambda \rightarrow 0$ if $a\neq 0$. But since there
 is no barrier to tunnel through the tunneling  condition will
simply imply that the universe stays at the origin $a=0$ and $ \Lambda$
is left undefined.

 One can seemingly obtain large or small $|\Lambda|$
depending on how one applies the boundary condition (the limits of
integration in eq.(4) ). When considering a -ve $\Lambda$ it
seems that we should conclude that the universe will
wish to stay at the origin and no predictions about $\Lambda$ should
be drawn from the factor $\exp{\sqrt{|\Lambda|}a^3}$.

 We generally  return to the case
of positive values of $\Lambda$ again from now;
but we keep in mind that the HH boundary condition which gives the
``wanted'' factor of $\exp(1/\Lambda)$ for positive $\Lambda$
also apparently  gives the prediction
of $\Lambda\rightarrow -\infty$ when $\Lambda$ is negative.

In this Euclidean region  we have found that the possible values of $\Lambda$
depend upon the choice of boundary conditions.
This point has also been mentioned  by Kiefer [14] in the context of wave
packet solutions to the WDW equation.
 Because the choice of boundary conditions
is not known a priori, it seems that to simply choose the boundary condition
that solves the cosmological constant problem is merely to pass  the
problem down the line.
 What is required is a measure of solutions to the WDW equation
which give either large or small $\Lambda$. Fortunately any
possible wave function
\begin{equation}
\Psi \sim \alpha\; \exp(1/\Lambda)+\beta\; \exp(-1/\Lambda)
\end{equation}
 ($\alpha, \beta $ arbitrary complex coefficients)
 will have the critical behaviour
at $\Lambda=0$,  even  for $\alpha$ small. It also
would make $D\rightarrow 0$  (defined in ref.[15]) and
according to the measure given in ref.[15]  a typical wavefunction
has $D<1$. This lends credence to the claim of Coleman that the
mechanism is somewhat immune from the choice of boundary conditions.
Although this wavefunction (11) will work alright for $+ve$ $\Lambda$ its
modified form cf. eq(8) for negative $\Lambda$ will be peaked at
$\Lambda\rightarrow -\infty$. The tunneling boundary
condition
predicts $\Lambda $ large for positive $\Lambda$, and also gives the
preferable
prediction that $\Lambda\rightarrow 0$ when it is negative. \\

 We consider next what happens when the universe starts
in a Lorentzian region where the WKB wave functions have the oscillating
behaviour $\Psi\sim \exp(\pm iS)$.
The exponents in the terms  $\exp(\pm iS)$ no longer have any critical
influence, but instead
the pre-factor contains any dominant behaviour.
We do however have to exclude the Euclidean regime from the
expression for the action cf. eq.(5) i.e. the lower
limit in the integral is taken to be $a=\Lambda^{-1/2}$. Otherwise
we would simply introduce the factors $\exp(\pm 1/\Lambda)$ again
and reach similar conclusions.\footnote {
 For the
same reason these factors
appear when you wish to normalize the wavefunction as $a\rightarrow 0$
-see e.g.[16]}

Typically the wavefunction has the form
\begin{equation}
\Psi \sim \frac{1}{a\sqrt{a^2\Lambda-1}}\left (e^{iS}+e^{-iS}\right )
\;\;.
\end{equation}
There is a similar peak around $a^2\Lambda\sim 1$ as the WDW potential is
zero. For $a$ fixed and $a^2\Lambda>>1$  then $\Psi^2\sim 1/\Lambda$ so
that larger values of $\Lambda$ are suppressed inversely.
 These are again
initial conditions to be followed by classical evolution,
and it would appear correct to assume the
quantum behavior made predictions for an initially small universe. The
initial value of $\Lambda $ would therefore appear large which would
produce an inflationary phase.

 This prediction occurs for both HH
and tunneling boundary conditions since they only determine
which combination of $\exp(iS)$ and $\exp(-iS)$ to take.
There is a heuristic reason to see this: in the Lorentzian
region the HH and tunneling solutions  look almost alike
( damped oscillations) and so should not differ much in their predictions.
Contrast this with their behaviour in the Euclidean regime - see eg.
Fig.(11.2) in ref. [7]. \footnote{ We  should
perhaps be careful and say the
prediction is not strongly dependent on boundary
conditions since there might be more unusual ways
of imposing them cf. Ref.[17]. Note the slight discrepancy with
Cline [17] who using  a Lorentzian path integral
approach concluded that $P(\Lambda)\sim 1$ so that any
$\Lambda$ is equally likely. In the Euclidean
region he found that
only special boundary conditions gave the
$\exp(1/\Lambda)$ factor.}

Similar predictions could be made if we considered a spatially
open $k=-1$ model together with a -ve $\Lambda$. This has Lorentzian
behaviour for small $a$ beyond which is a Euclidean regime. The peak
would be around $a^2|\Lambda |\sim -k$. Notice  how the
spatial curvature is crucial for any predictions about $\Lambda$ . If we
set $k=0$ then $P(\Lambda) \sim 1/(a^4\Lambda)$ and we would obtain the
prediction that $\Lambda \rightarrow 0$, although without
the exponential peak.

 It might appear that this property $a^2\Lambda \sim 1$
 is an artifact of using WKB solutions which are
simply blowing up at the  turning point  $U\equiv a^2-\Lambda a^4=0$.
However exact solutions of the WDW equation can be found and this behaviour
remains. For example the equation [18]
\begin{equation}
\left (\frac{d^2}{da^2}+\frac{p}{a}\frac{d}{da}-U\right )\Psi(a)=0
\end{equation}
with $p$ a factor ordering correction, has solutions
\begin{equation}
\Psi\sim a^{\frac{1-p}{2}}\left \{J_{(1-p)/4}(\sqrt{-U}a)+
Y_{(1-p)/4}(\sqrt {-U}a)\right \}
\end{equation}
Since both Bessel functions $J_{\nu}(x)$ and $Y_{\nu}(x)$
 both decrease for increasing $x$ they
both take their maximum value when $x=0$ and as we require $a>0$ this
occurs for $U=0$ , so again when $\Lambda \sim 1/a^2$. For $p\neq 1$ this
is slightly modified $\sqrt {-Ua}\sim small$.

The addition of additional matter fields is likely to round off this
spike at $\Lambda \sim 1/a^2$ cf. ref.[19].
We see that the initial value of $\Lambda$ is expected to be large in
the Lorentzian regime provided $a$ is not large in Planck units. If the
initial size of the universe is `big' $\sim 1\;cm$ the
probable value of $\Lambda$ is
smaller but still huge compared to its present value cf. Ref.[11].

Let us finally try to understand the Euclidean regime results
(when $U\geq0$)
in terms of the
solutions  eq.(13). For factor ordering $p=1$ the solutions simplify
to
\begin{equation}
\Psi\sim K_0(aU^{1/2})+I_0(aU^{1/2})
\end{equation}
Since $K_0(x)\rightarrow \infty $ as $x\rightarrow 0$ it picks out
the $U=0$ or $\Lambda\sim 1/a^2$ case. The other Bessel function
$I$ increases with increasing $aU^{1/2}\equiv a^2(1-\Lambda a^2)
^{1/2}$ so is maximized for $\Lambda=0$,
or  negative $\Lambda$ if we allow it. This is as expected since the
tunneling boundary condition is the decaying solution $K$ and the
HH one a mixture of both $I$ and $K$- see eg.ref.[20].

Recently a WDW equation corresponding to a classical signature change
has been obtained [21]. Such an approach always has an oscillating
wavefunction and so does not have the possibility of having
Baum-Hawking factors $\exp(1/\Lambda)$.  It seems only
consistent with  tunneling boundary
conditions [21] and for this reason we suspect it
would also predict $\Lambda$ large when $\Lambda$ is a dynamical variable.\\

{\bf Conclusions}

In the Euclidean regime the initial value of $\Lambda$ is expected
large if tunneling boundary conditions or tunneling like behaviour
is correct. This would not be suitable for setting $\Lambda$ small
but would allow for an Inflation regime to proceed. If
we consider HH boundary conditions then
you can get the factor $\exp(1/\Lambda)$ (or if $\Lambda$ is -ve the
anomalous
$\exp{\sqrt{|\Lambda |}a^3}$ factor).
 The two behaviours are complimentary:
 we could not have Inflation together with
$\Lambda \rightarrow 0$ unless some other dynamical mechanism could
give the two mechanisms differing time scales cf. Ref.[22].

 In fact, if
the tunneling boundary condition is correct and $\Lambda$ correspondingly
large, we would not want the wormhole mechanism to take place. This would
have the effect of transferring
 the large value of $\Lambda$ in the average ``quantum foam
of universes'' to
any universe, so giving a large cosmological constant in
our universe.\\

We then considered  the purely  Lorentzian regime and found that the
predictions in this case are
 not dependent on the boundary
conditions.
\footnote{In this regard we essentially agree with the analysis of Strominger
[11], except for his conclusion that this implies $\Lambda \rightarrow
\simeq 0$.}
 We found an initial value of $\Lambda \sim k/a^2$
 and  since we expect the universe to start small, due to
quantum gravity processes, the corresponding value
of $\Lambda$ is large.
We are still left with the problem of why $\Lambda$ should be a
dynamical variable with a distribution function. The 3-form
(axion) field might still work in this regard even if
 its wormhole solution cannot be invoked cf.ref.[2].\\
\\
{\bf Acknowledgement}\\
I should like to thank A. Carlini and D. Solomons for helpful
comments.\\

{\bf References}\\
\begin{enumerate}
\item S. Coleman, Nucl. Phys. B 310 (1988) p.643.
\item S. Weinberg, Rev. of Mod. Phys. 61 (1989) p.1.
\item E. Baum, Phys. Lett. 133 B (1983) p.185.
\item S.W. Hawking, Phys. Lett. 134 B (1983) p.403.
\item J.J. Halliwell, in {\em Quantum Cosmology and Baby Universes}
eds. S. Coleman et al. (World Scientific, Singapore) 1991.
\item A.D. Linde, {\em Particle  Physics and Inflationary Cosmology}
(Harwood, Switzerland 1990).
\item E.W. Kolb and M.S. Turner, {\em The Early Universe}, (Addison-
Wesley, USA 1990).
\item A. Vilenkin, Phys. Rev. D 30 (1984) p.549\\
A.D. Linde, Sov. Phys. JEPT 60 (1984) p.211\\
V.A. Rubakov, Phys. Lett. 148 B (1984) p.280.\\
Y.B. Zeldovich and A.A. Starobinski, Sov. Astron. Lett. (1984) p.135.
\item A. Vilenkin, Phys. Rev. D 37 (1988) p.888.
\item V.A. Rubakov and P.G. Tinyakov, Nucl. Phys. B 342 (1990) p.430.\\
 G. Lavrelashrili, V.A. Rubakov and P.G. Tinyakov, in {\em
Gravitation and Quantum Cosmology} eds. A. Zichichi et al. (Plenum
Press, New York, 1991) p.87.
\item A. Strominger, Nucl. Phys. B 319 (1989) p.722.
\item J.B. Hartle and S.W. Hawking, Phys. Rev. D 28 (1983) p.2960.
\item A. Vilenkin, {\em Approaches to quantum cosmology} preprint (1994).
\item C. Kiefer, Nucl. Phys. B 341 (1990) p. 273.
\item G.W. Gibbons and L.P. Grishchuk, Nucl. Phys. B 313 (1989) p.736.\\
 L.P. Grishchuk and Y.V. Sidorov, Sov. Phys. JEPT 67 (1988) p.1533.
\item E. Fahri, Phys. Lett. B 219 (1989) p.403.
\item J. M. Cline, Phys. Lett. 224 B (1989) p.53.
\item D.H. Coule, Class. Quan. Grav. 9 (1992) p.2353.
\item T. Banks, Nucl. Phys. B 309 (1988) p.493.
\item D.N. Page, in {\em Proceedings of Banff summer research institute
on Gravitation} eds. R. Mann and P. Wesson (World Scientific, Singapore)
1991.
\item J. Martin, Phys. Rev. D 49 (1994) p.5105.
\item T. Fukuyama and M. Morikawa, Class. Quan. Grav. 7 (1990) p.823.
\end{enumerate}

{\bf Figures}\\
Fig.1\\
The Wheeler-DeWitt potential $U$. The Euclidean regime has $U\geq 0$
beyond which it is Lorentzian. The tunneling boundary condition
describes the decay from the origin to $a=\Lambda^{-1/2}$.
\end{document}